\documentclass[12pt,preprint]{aastex}
\usepackage{natbib}
\usepackage{multirow}
\shorttitle{Formation of Propellers in Dense Rings}
\shortauthors{Michikoshi \& Kokubo}

\begin{document}

\title{
Formation of a Propeller Structure by a Moonlet in a Dense Planetary Ring
}
\author{
Shugo Michikoshi\altaffilmark{1} and 
Eiichiro Kokubo\altaffilmark{1,2}
}
\altaffiltext{1}{
Center for Computational Astrophysics, National Astronomical Observatory of Japan, Osawa, Mitaka, Tokyo 181-8588, Japan
}
\altaffiltext{2}{
Division of Theoretical Astronomy, National Astronomical Observatory of Japan, Osawa, Mitaka, Tokyo 181-8588, Japan
}
\email{michikoshi@cfca.jp and kokubo@th.nao.ac.jp}

\begin{abstract}

The Cassini spacecraft discovered a propeller-shaped structure in
 Saturn's A ring.
This propeller structure is thought to be formed by gravitational
 scattering of ring particles by an unseen embedded moonlet.
Self-gravity wakes are prevalent in dense rings due to gravitational
 instability. 
Strong gravitational wakes affect the propeller structure.
Here, we derive the condition for formation of a propeller structure by a
 moonlet embedded in a dense ring with gravitational wakes. We find that a propeller structure is formed when the wavelength of the gravitational wakes is smaller than the Hill radius
 of the moonlet.
We confirm this formation condition by performing numerical simulations.
This condition is consistent with observations of propeller
 structures in Saturn's A ring. 

\end{abstract}

\section{Introduction \label{sec:intro}}

A moonlet embedded in a planetary ring tends to open a gap, such as a
 Keeler or Encke gap, due to gravitational scattering
 \citep[e.g.,][]{Lissauer1981}.  
Conversely, viscous diffusion of ring particles tends to close a gap. 
If a moonlet is sufficiently large, it will form a fully circular gap, whereas 
a small moonlet will form only a partial gap that consists of two
 azimuthally aligned lobes shaped like a propeller (hereafter, we refer to this structure as a propeller).
Using the viscous fluid model, \cite{Spahn2000} and \cite{Sremcevic2002}
 predicted the formation of propellers by small moonlets.
The Cassini spacecraft discovered propellers in Saturn's A ring
 \citep{Tiscareno2006}.  
The moonlets are so small that they cannot be directly detected.
Since the radial separation between the two lobes is related to the Hill
 radius of a moonlet, the size of a moonlet with a propeller can be
 estimated. 
Most of the known propellers are concentrated within narrow bands in
 the A ring. 
Moonlets with propellers have radii ranging from tens of meters to
 a kilometer \citep{Tiscareno2010}. 
The size distribution of moonlets has a steeper power-law index
 than that of ring particles \citep{Tiscareno2008}.

An $N$-body simulation is a powerful tool for studying ring dynamics where
 collisions and gravitational interactions play important roles. 
\cite{Seiss2005} and \cite{Lewis2009} performed $N$-body simulations of 
 propeller formations. 
\cite{Seiss2005} considered collisions between particles but did not take into account of the self-gravity of particles. 
They treated the effect of self-gravity as increasing the
 vertical frequency \citep{Wisdom1988}.
They confirmed the formation of propellers and investigated the scaling
 law of the propeller size discussed in \cite{Sremcevic2002}.
\cite{Lewis2009} included self-gravity and the size distribution
 of particles.
They adopted a low optical depth $\tau=0.1-0.2$, which is smaller than
 that of the A ring. 
Self-gravity causes spontaneous formation of gravitational wakes
 \citep[e.g.,][]{Salo1995, Daisaka1999}. 
They showed that self-gravity with a size distribution destroys
 the propeller structure when the ratio of the moonlet mass to the upper
 limit of the mass distribution is less than 30.

In this study, we perform real-scale $N$-body simulations of a
 small moonlet embedded in dense rings.
The optical depth of the A ring is as high as about $0.3$--$0.5$, while
 that of the B ring is larger than unity. 
Distinct and large gravitational
 wakes form in rings with large optical depths
\citep[e.g.,][]{Salo1995,Daisaka1999}. 
Such large gravitational wakes may alter the structures around the
 embedded moonlet.
We investigate the condition for propeller formation in a dense ring
 in which gravitational wakes are prevalent.
Section 2 summarizes the calculation method.
In Section 3, we present the simulation results and derive the
 condition for propeller formation.
Section 4 consists of a discussion and a summary.

\section{Calculation Method \label{sec:simultaion}}

We consider a small computational domain of a ring such that 
 $L_x, L_y \ll a$, where $L_x$ and $L_y$ are the width and length of the
 computational domain and $a$ is the distance from Saturn.
We introduce a local Cartesian coordinate system ($x,y,z$). 
The origin revolves around Saturn with the Kepler angular velocity
 $\Omega$ and is located at the center of the computational domain. 
The $x$-axis is directed radially outward, the $y$-axis is parallel to the direction
 of rotation, and the $z$-axis is normal to the $x$--$y$ plane. 
The computational domain has periodic boundary conditions and is
 surrounded by eight copies \citep[e.g.,][]{Wisdom1988, Salo1995}.
The equation of motion is linearized, which is referred to as the Hill
 equation \citep[e.g.,][]{Petit1986,Nakazawa1988}. 
The equation of motion is integrated using a second-order leapfrog
 scheme with a variable time step.
The sizes of the computational domain are $L_x = 10 r_\mathrm{H}$ and
 $L_y=80 r_\mathrm{H}$, where $r_\mathrm{H}$ is the Hill radius of the
 moonlet given by $r_\mathrm{H} = (M/3M_\mathrm{s})^{1/3} a$, where $M$
 and $M_s$ are the masses of the moonlet and Saturn, respectively.

We consider the self-gravity of ring particles and calculate it by
 directly summing the gravitational interactions of all pairs.
We calculate not only the gravitational interactions inside the domain
 but also those from the surrounding copies \citep{Salo1995}. 
The gravitational interactions, which are the most computationally expensive part of
 $N$-body simulations, are calculated using a programmable, special-purpose
 computer, GRAPE-DR \citep{Makino2007}.

The frictionless impact model for hard spheres is adopted
 \citep[e.g.,][]{Richardson1994,Daisaka1999}. 
In a collision, the normal component of the relative velocity is reduced 
 by a factor $\epsilon$, which is the restitution coefficient in the
 normal direction. 
The tangential component of the relative velocity is conserved.
We adopt a restitution coefficient model that was determined by a
 laboratory experiment. 
The normal restitution coefficient is \citep{Bridges1984}
\begin{equation}
 \epsilon = 0.34 \min 
  \left(\left(\frac{v_\mathrm{n}}{1\mathrm{cm}/\mathrm{s}}\right)^{-0.234},
   1 \right), 
\end{equation}
 where $v_\mathrm{n}$ is the normal impact velocity.

We consider the size distribution of particles. 
We adopt a power-law model $n dR= (R/R_0)^{-q} dR$ for
 $R_\mathrm{min} < R < R_\mathrm{max}$, where $R$ is the radius of particles,
 $R_0$ is the radius for normalization, $q$ is the power-law index for
 the distribution, and $R_\mathrm{min}$ and $R_\mathrm{max}$ are the
 minimum and maximum sizes of particles, respectively. 
We adopt $q=2.8$ \citep{Zebker1985}.
The density of particles is $0.5 \,\mathrm{g}/\mathrm{cm}^{3}$.
The initial Toomre parameter $Q=\Omega c/3.36 G \Sigma$ is set to $Q=2$ \citep{Toomre1964}, where $c$ is the velocity dispersion of ring particles.

The moonlet is fixed at the origin of the coordinates. 
We assume that the moonlet radius is $150 \, \mathrm{m}$ and 
 its bulk density is $0.9\, \mathrm{g}/\mathrm{cm}^3$.
The semimajor axis and the Hill radius of the moonlet are a=117000 km and $r_\mathrm{H} = 228 \,\mathrm{m}$, repectively.
This semimajor axis corresponds to the outer region of the B ring.

\section{Results \label{sec:result}}

\subsection{Formation and Non-Formation of Propellers}

We demonstrate formation and non-formation of propellers with low and high surface density ring
 models.
The typical minimum surface density in the B ring is
 estimated to be $240$--$480\,\mathrm{g}/\mathrm{cm}^2$ \citep{Robbins2010},
 while the surface density at its outer edge is estimated to be
  $30$--$70\, \mathrm{g}/\mathrm{cm}^2$ \citep{Spitale2010}.
We adopt $\Sigma_0 = 62\, \mathrm{g}/\mathrm{cm}^2$ and
 $414\, \mathrm{g}/\mathrm{cm}^2$ for the low and high surface density 
 models, respectively.
The size range of particles is inferred from stellar occultation as
 $R_\mathrm{min} \simeq 30\,\mathrm{cm}$ and
 $R_\mathrm{max} \simeq 20\, \mathrm{m}$ \citep{French2000}.
However, due to the limitations of the available computing resources, we have to
 adopt the larger and smaller values respectively for $R_\mathrm{min}$ and
 $R_\mathrm{max}$ of $R_\mathrm{min} = 2\, \mathrm{m}$ and 
 $R_\mathrm{max} = 10\, \mathrm{m}$ to fit the surface density to the
 value inferred from the density wave \citep[e.g.,][]{Tiscareno2007}.
This small size range gives somewhat unrealistic dynamical optical
 depths of $\tau_0 = 0.18$ and 1.2 for the low and high surface density
 models, respectively.
However, the surface density controls the basic dynamics of
 dense rings in which gravitational wakes develop.
As is shown later, the condition for propeller formation
 depends on the surface density and is independent of the optical
 depth. 

\begin{figure*}
\plotone{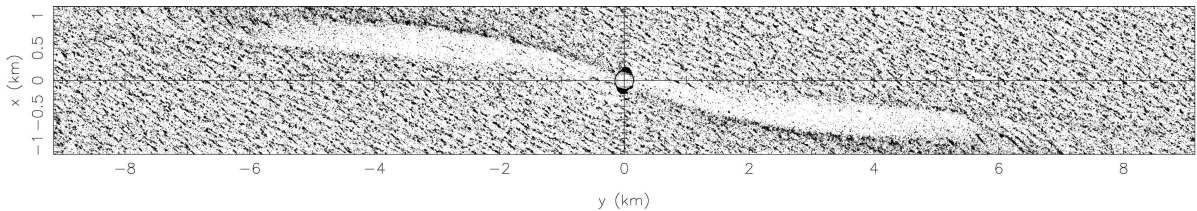}
\caption{
Snapshot of low surface density model at $t=1.0T_\mathrm{K}$. 
The initial surface density is $\Sigma_0=62 \mathrm{g}/\mathrm{cm}^2$. 
The circle at the center of the computational box is the moonlet. 
A propeller structure is visible around the moonlet.
}
\label{fig:tau01}
\end{figure*}

First, we show formation of a propeller in the low surface density
 model. 
Figure \ref{fig:tau01} shows a snapshot of the low surface density model at $t=1.0 T_\mathrm{K}$, where $T_\mathrm{K}$ is the Keplerian period $2 \pi / \Omega$.
A propeller-shaped feature is clearly visible in the weak
 gravitational wakes. 
The surface density decreases considerably in the two lobes
downstream of the moonlet.
They are aligned in the orbital direction and are symmetric about the
 moonlet.
The minimum density in the propeller is located at about 
 $x=\pm 2 r_\mathrm{H} = 456\, \mathrm{m}$ and $y=\mp 4\, \mathrm{km}$ 
 \citep{Seiss2005}. 
The radial separation between the two lobes is about $4r_\mathrm{H}$ and the radial width of a
 single lobe is about $2r_\mathrm{H}$ 
 \citep{Spahn2000, Sremcevic2002, Seiss2005, Tiscareno2008, Lewis2009}.
The length of the lobe in the azimuthal direction is about $5\, \mathrm{km}$. 
We define the propeller region as the region of $-3<x/r_\mathrm{H}<-2$
 and $15<y/r_\mathrm{H}<20$, which is the typical region of a propeller
 lobe \citep{Spahn2000, Sremcevic2002, Seiss2005, Lewis2009}.
The ratio of the time-averaged surface density in the propeller region to
 the initial surface density $\bar \Sigma / \Sigma_0$ is $0.17$. 

\begin{figure*}
\plotone{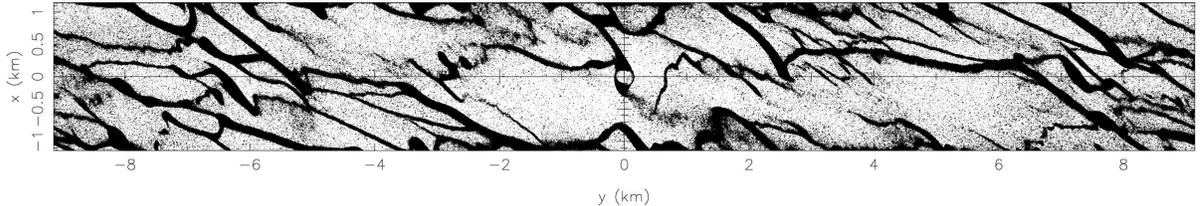}
\caption{
Same as for Figure \ref{fig:tau01}, but for the high surface density model with
 $\Sigma=414$ $\mathrm{g}/\mathrm{cm}^2$ at $t= 4.0T_\mathrm{K}$. 
Strong gravitational wakes form and a propeller structure is not clearly observed. 
}
\label{fig:tau10}
\end{figure*}

Next, we demonstrate non-formation of a propeller in the high surface
 density model.
Numerical simulations reveal that a strong wake structure due to
 gravitational instability forms for this parameter
 \citep{Salo1995,Daisaka1999}.
As expected, until $t=0.5 T_\mathrm{K}$, gravitational instability
 occurs and wakes start forming.
Figure \ref{fig:tau10} shows a snapshot of the high surface density
 model at $t=4.0 T_\mathrm{K}$. 
Strong gravitational wakes form and no propeller structure is 
 clearly observed.
The ratio of the time-averaged surface density in the propeller region to
 the initial surface density $\bar \Sigma / \Sigma_0$ is $0.41$. 
The gravitational wakes seem to be almost unaffected by the moonlet.
The typical distance between wakes is approximated by the critical
 wavelength of gravitational instability 
 $\lambda_\mathrm{cr} = 4 \pi^2 G \Sigma / \Omega^2$, where $\Sigma$ is
 the ring surface density \citep[e.g.,][]{Julian1966,Salo1995}.
In the high surface density model, the typical distance between wakes
 is $\lambda_\mathrm{cr} = 459$ m, which is larger than the Hill radius
 of the moonlet.

\subsection{Condition for Propeller Formation \label{sec:conf}}

The numerical simulation results indicate that
 propeller formation depends on the surface density. 
Below, we derive the formation condition for propellers and confirm its validity by performing $N$-body simulations.

The clumps in gravitational wakes typically have a mass of $\sim
 \Sigma \lambda_\mathrm{cr}^2$. 
If the clump mass is greater than the moonlet mass $M$, the 
 gravitational wakes may not be affected by gravitational scattering
 due to the moonlet. 
Comparing the clump mass with the moonlet mass, we obtain the following condition
 for propeller formation
\begin{equation}
 \lambda_\mathrm{cr} \lesssim r_{\rm H}.
\label{eq:condition}
\end{equation}
Note that this condition can also be derived by comparing the typical velocity due to scattering by the moonlet with the velocity dispersion of ring particles \citep{Lewis2009}. 
If the velocity dispersion of particles exceeds the Hill velocity $r_\mathrm{H} \Omega$, which is a typical velocity due to scattering by the moonlet, the gravitational wake should be almost unaffected by the moonlet.
The velocity dispersion of ring particles $c$ is determined by
 $Q=2$ \citep{Salo1995, Daisaka1999}.
Thus, the ratio of $c$ to $r_\mathrm{H} \Omega$ is about 
 $\lambda_\mathrm{cr} / r_\mathrm{H}$. 

Using the moonlet radius and the ring surface density, Equation
 (\ref{eq:condition}) can be rewritten as
\begin{eqnarray}
\Sigma < \Sigma_\mathrm{cr} 
 & \equiv 
 & C \left( \frac{M_\mathrm{s}^2 \rho R^3}{144 \pi^5 a^6} \right)^{1/3} 
 \nonumber  \\  
 & = 
 & 167 \, \mathrm{g}/\mathrm{cm^2} \left(\frac{C}{1.5}\right)
 \left(\frac{a}{1.3 \times 10^5 \, \mathrm{km}}\right)^{-2}
 \left(\frac{\rho}{0.9 \, \mathrm{g}/\mathrm{cm}^3}\right)^{1/3}
 \left(\frac{R}{100 \, \mathrm{m}}\right)
\label{eq:condition2}
\end{eqnarray}
 where $C$ is the non-dimensional constant of order unity (see
 discussion below) and $\rho$ and $R$ are respectively the density and radius of the
 moonlet.

To check the validity of this condition, we perform $N$-body simulations
 for various $\Sigma$ and $R$.  
To investigate a wide range of $\Sigma$, we adopt equal-sized (10
 m) ring particles, which reduces the number of particles.
The density of particles is $0.5 \mathrm{g}/ \mathrm{cm}^3$.
We assume that the restitution coefficient is constant at $\epsilon=0.1$.
The moonlet density is $0.7 \mathrm{g}/ \mathrm{cm}^3$.

\begin{figure}
\plotone{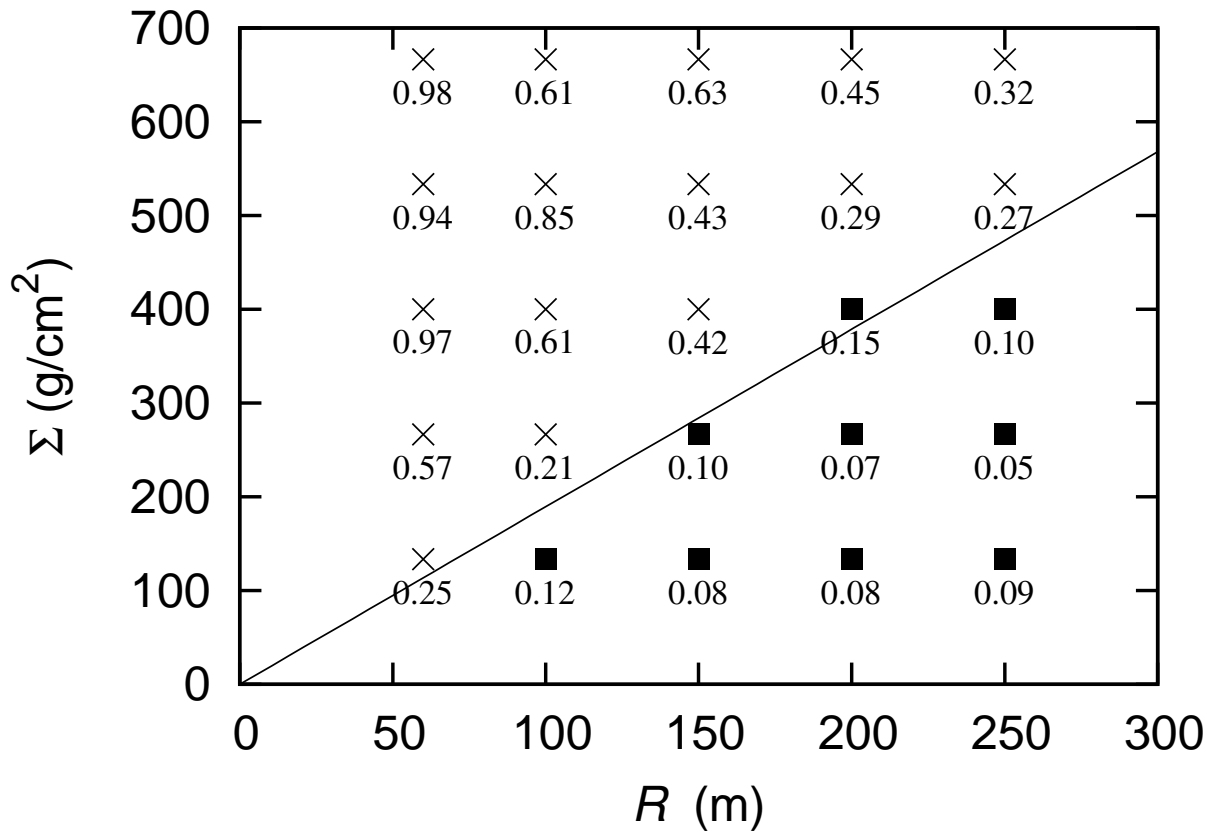}
\caption{
Condition for propeller formation in Saturn's B ring in the $R$-$\Sigma$ plane.
Filled squares denote models in which clear propellers form and
 crosses denote models in which no propellers form. 
We show the ratio of the time-averaged surface density in the propeller region to the initial surface density $\bar \Sigma / \Sigma_0$ at each point.
The solid line indicates the condition for propeller formation 
 estimated by Equation (\ref{eq:condition2}) with $C=1.68$.
}
\label{fig:summary.eps}
\end{figure}

We perform 25 simulations with different ring surface densities and
 moonlet radii.
We vary the moonlet radius from $60\,\mathrm{m}$ to $250 \,\mathrm{m}$ and
 the surface density from $\Sigma = 133\, \mathrm{g}/\mathrm{cm}^2$ to
 $\Sigma = 670 \,\mathrm{g}/\mathrm{cm}^2$. 
The ratio of the radius to the Hill radius is 0.64.
The simulation time is $6 T_\mathrm{K}$.
We consider
 that a propeller forms if $\bar \Sigma / \Sigma_0 < 0.2$ in the propeller region.
We confirmed that a propeller-shaped structure is clearly observed
 when this condition is satisfied.
If this condition is not satisfied, 
no distinct steady propeller structure is observed (although the ring particles are
 affected by the moonlet to some extent).
Figure \ref{fig:summary.eps} summarizes the results. It 
 clearly shows that Equation (\ref{eq:condition2}) is consistent with the
 simulation results. 
We find that $C=1.68$ explains the $\bar \Sigma / \Sigma_0 = 0.2$ boundary well.
We also find that $C$ is approximately proportional to the boundary value of $\bar \Sigma / \Sigma_0$.

For a moonlet with a radius of $150\, \mathrm{m}$ and a density of
 $0.9\, \mathrm{g}/\mathrm{cm}^3$, the critical surface density is $308\,
 \mathrm{g/cm^2}$. 
In the low surface density model, the surface density is $62\,
 \mathrm{g/cm^2}$, which is smaller than the critical value. 
In this case, a propeller is clearly observed.
On the other hand, in the high surface density model, since the surface
 density is $414 \mathrm{g/cm^2}$, which is greater than the critical value,
 a gravitational wake prevails over a propeller.
Consequently, no propeller is observed.

\begin{figure}
\plotone{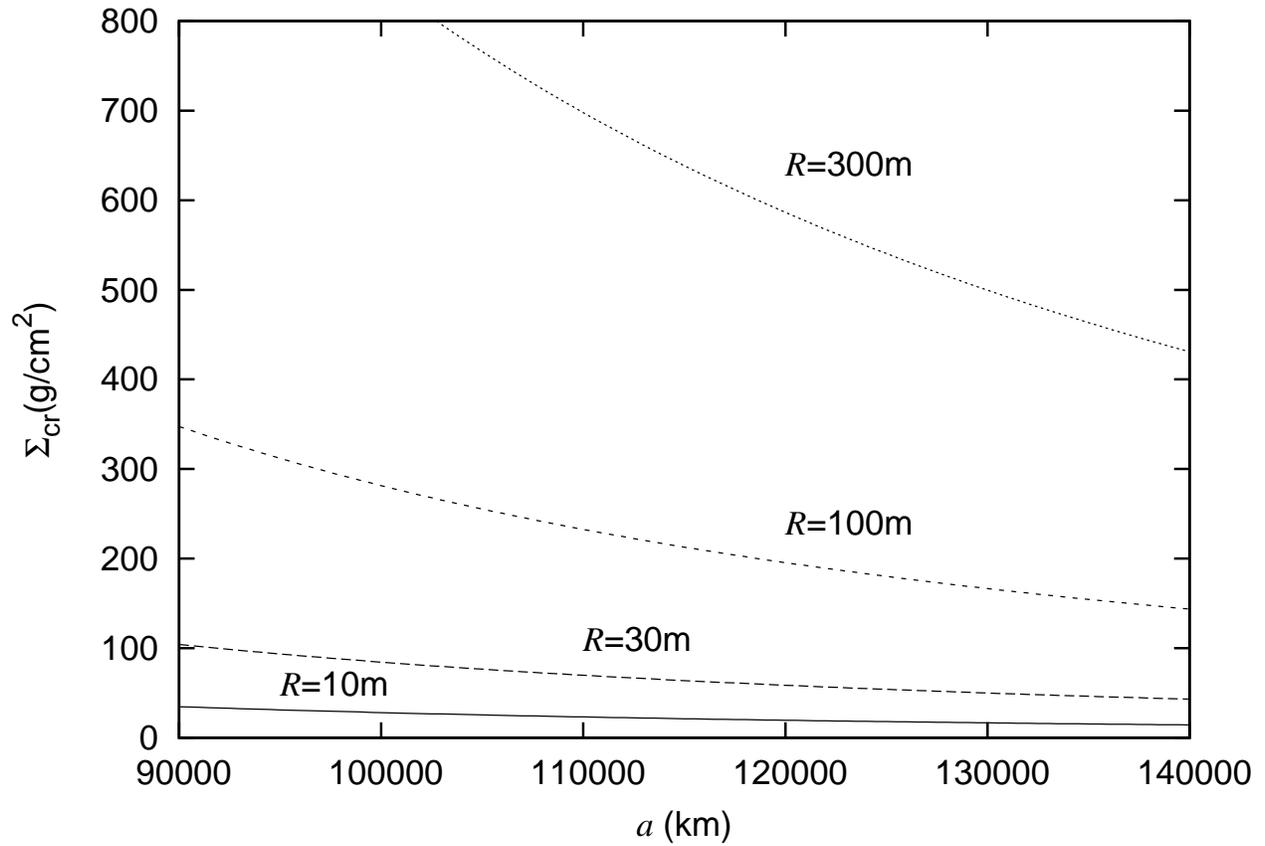}
\caption{Critical surface density as a function of distance from Saturn for moonlets with 
$R=10\, \mathrm{m}$ (solid curve), $30\, \mathrm{m}$ (dashed curve),
 $100\, \mathrm{m}$ (short dashed curve), and $300\, \mathrm{m}$ (dotted
 curve). 
The moonlet density is $0.9\, \mathrm{g}/\mathrm{cm}^3$.
}
\label{fig:a-Sigma}
\end{figure}

Figure~\ref{fig:a-Sigma} shows the critical surface density plotted as a function of 
 distance from Saturn for moonlets with
 $\rho = 0.9\, \mathrm{g}/\mathrm{cm}^3$ and $R = 10$, 30, 100, and 300 m. 
Since $\lambda_\mathrm{cr}$ increases with $a$ faster than $r_\mathrm{H}$, 
 $\Sigma_\mathrm{cr}$ decreases with increasing $a$.
The typical surface density in the A ring is about
 $40 \mathrm{g} / \mathrm{cm}^2$ 
 \citep[e.g.,][]{Esposito1983, Tiscareno2008}.
Assuming $\rho=0.9 \mathrm{g}/\mathrm{cm}^3$, we obtain the critical
 moonlet radius for propeller formation to be $20\, \mathrm{m}$. 
Known moonlets with propellers have radii ranging from
 $20\, \mathrm{m}$ to a kilometer \citep{Tiscareno2008, Tiscareno2010}.
Equation (\ref{eq:condition2}) is consistent with this observational data.

Note that this condition is not applicable to rings with low
 surface densities in which gravitational wakes do not develop well.
In this case, gravitational scattering of individual particles is
 important and the condition discussed in \cite{Lewis2009} should be
 applicable, namely that propeller formation is controlled by the ratio of the maximum ring particle mass to the moonlet mass.

\section{Summary and Discussion}

We have performed local $N$-body simulations to investigate the
 formation of a propeller by a moonlet.
By performing real-scale simulations, we demonstrated that in the B ring a
 moonlet with radius $R = 150\,\mathrm{m} $ forms a propeller in a low
 surface density ring ($60 \,\mathrm{g}/\mathrm{cm}^2$), whereas it does
 not in a high surface density ring ($414 \,\mathrm{g}/\mathrm{cm}^2$)
 in which the ring dynamics are dominated by gravitational wakes.
These results indicate that propeller formation depends on the surface
 density.
Comparing the moonlet mass with the typical mass of a gravitational
 wake, we derived the condition for propeller formation that the
 characteristic length of the gravitational wakes given by the
 critical wavelength for gravitational instability be shorter than the 
 Hill radius of the moonlet, $\lambda_\mathrm{cr} \lesssim r_\mathrm{H}$.
 We confirmed this by $N$-body simulations.
In a ring with gravitational wakes, the characteristic length of
 wakes is more important than the size of individual ring particles
 since a wake is a coherent structure. 
Our condition is consistent with observations of propellers in
 Saturn's A ring which revealed observational signatures of gravitational
 wakes \citep[e.g.,][]{French2007}.

The Cassini spacecraft recently discovered a new putative ``moonlet'' in the B ring (S/2009 S1) \citep{Porco2009}. 
The diameter of S/2009 S1 is inferred to be 300 m if it is orbiting in
 the same plane as the ring particles.
Its radial distance from the center of Saturn is 116,914 km, which is at
 the outer edge of the B ring \citep{Spitale2010}.
Surprisingly, no propeller was observed around S/2009 S1.
For S/2009 S1, the critical surface density for propeller formation is
 estimated to be $\Sigma_\mathrm{cr} = 308 \mathrm{g/cm^2}$, which is
 higher than the typical surface density of the outer edge of the B
 ring. 
This suggests two possibilities: the surface density around
 S/2009 S1 is locally high enough to prevent propeller formation or
 S/2009 S1 is not a moonlet but a transient feature such as meteoroid
 impact or temporary clumps.
Further observation is necessary to determine the nature of S/2009 S1.

For simplicity, we assumed that the moonlet is fixed.
If the moonlet can move, the moonlet will be scattered by gravitational
 wakes, which leads to stochastic change in the moonlet orbit
 \citep{Lewis2009}. 
This random motion of the moonlet may hinder propeller formation.
It is important to determine the shape of propeller structures to interpret 
 observations.
The shape may depend on various ring parameters such as the ring viscosity
 and the size distribution of ring particles. 
We intend to investigate these problems in a subsequent study.

Observations of propeller structures reveal bright regions rather than
 low-density features such as a propeller gap \citep{Sremcevic2007}.
However, the implied density enhancement near the moonlet has not
 been observed in numerical simulations \citep{Lewis2009}. 
One possible explanation for the brightness enhancement is
 collisional debris released from ring particles \citep{Sremcevic2007}.
Gravitational scattering of ring particles by the moonlet increases the
 impact velocity between ring particles.
Thus, debris may be released from ring particles in the vicinity of
 the moonlet in high-velocity collisions, which increases the reflectivity.
As shown in Section \ref{sec:conf}, the condition for propeller gap formation is equivalent to that the Hill velocity of the moonlet $r_\mathrm{H} \Omega$ is higher than the velocity dispersion of ring particles.
Therefore, it may be applicable to the condition for local debris
 enhancement.
We intend to study this in the future.

We wish to thank Joseph A. Burns for stimulating discussions on the
 observations of propellers by the Cassini spacecraft.
E. K. is grateful to the Isaac Newton Institute for Mathematical
 Sciences for its hospitality during his visit at the beginning of this study.
The numerical calculations were performed on the GRAPE system at the
 Center for Computational Astrophysics of the National Astronomical
 Observatory of Japan.

\end{document}